# Predicting electronic stopping powers using stacking ensemble machine learning method

Fatemeh Akbari*, Somayeh Taghizadeh, Diana Shvydka, Nicholas Niven Sperling, E. Ishmael Parsai

Department of Radiation Oncology, University of Toledo Health Science Campus, Toledo, OH, 43614 USA
*fatemeh.akbari@rockets.utoledo.edu

## Abstract

**Background:** Accurate electronic stopping power data is crucial for calculations of radiation-induced effects in a wide range of applications, from dosimetry and radiotherapy to particle physics. The data is dependent on the parameters of both the incident charged particle and the stopping medium. The existent Bethe theory can be used to calculate the stopping power of high-energy ions, but fails at lower energies, leaving incomplete and even contradictory experimental data, often expanded through extrapolations with fitting formula, as the only accessible resource. Moreover, the majority of the experimental data is available for elements only, further limiting the validity of fitting approaches for complex material compositions. A relatively novel machine learning methodology has been proven to be effective for exactly these types of problems.

**Purpose:** In this study, Stacking Ensemble Machine Learning (EML) algorithm was developed to predict electronic stopping power for any incident ion and target combination over a wide range of ion energies. For this purpose, five ML models, namely BR, XGB, AdB, GB, and RF, were selected as base and meta learners to construct the final Stacking EML.

**Methods:** 40,044 experimental measurements, from 1928 to the present, available on the International Atomic Energy Agency (IAEA) website were used to train machine learning (ML) algorithms. This database consists of 593 various ion-target combinations across the energy range of 0.037 to 985 MeV. For model training, the eleven most important features were selected. The model evaluation was performed using several error metrics, including R-squared ($R^2$), root-mean-squared-error (RMSE), mean-absolute-error (MAE), and mean-absolute-percentage-error (MAPE), on both the training and test datasets.

**Results:** Based on model performance evaluation tests, a stack of eXtreme Gradient Boosting (XGB) and Random Forest (RF) via Bagging Regressor (BR) meta-learner had the highest lowest error margin. The value of $R^2=0.9985$ indicated a near-ideal fit to all samples in the training data across the entire range of stopping powers. $R^2$=0.9955 for predictions made by the model on the unseen test data suggested that the model accurately predicted the test data.

**Conclusions:** The developed model resulted in highly accurate predictions for any ion-target combination across the whole particle energy spectrum. The associated model can serve as a universal tool to generate the stopping power data in a wide range of cases, regardless of the availability of experimental data or reliable theoretical equations. Overall, the demonstrated success of the developed tool testified to the power of machine learning approaches, and the suitability of the chosen models for solutions to practically important physics problems.

**Key words:** Stopping power, Machine learning, Stacking ensemble methods

# 1- Introduction

Characterization of the energy deposition in the process of charged particles moving through materials has attracted the interest of many researchers in a variety of fields, including fundamental particle physics, nuclear physics, radiation biology, and radiation physics [1, 2]. In medical physics stopping power parameter provides essential information for dosimetry purposes. Stopping power and absorbed radiation dose are strongly correlated and contribute to determining the radiation effects on living organisms and various human body tissues. The stopping power is defined as the average energy dissipation $E$ per unit path length $x$ of the moving charged particle, $(dE/dx)$; it depends on the charge and velocity of the incident ion and target material (composition, phase, density, etc.). The total stopping power is generally divided into three components [3]

$$S_{total}(E) = S_{el}(E) + S_{nuc}(E) + S_{rad}(E)$$

where $S_{el}$, is electronic stopping power caused by inelastic Coulomb collisions with an orbital electron of the medium atoms and $S_{nuc}$ is the nuclear term arising from elastic Coulomb collisions with the target nuclei. The radiative stopping power, $S_{rad}$, is due to the emission of bremsstrahlung produced by incident charged particle when interacting with the electric field of atomic nuclei. At low energies and/or when the ion is massive, radiative and nuclear components are negligible and total energy loss is characterized by electronic stopping power. Dividing stopping power by medium density yields a more useful quantity known as mass stopping power which is expressed in the unit of MeV·cm$^2$·g$^{-1}$.

Many experiments have been performed to measure the stopping power over the past few decades [4]. Experimental observations are important because they provide a benchmark for evaluating analytical studies and simulations. However, they are restricted to specific energy ranges and ion-target combinations, as studies with arbitrary ions and target materials are not practical. The most widely accepted model to obtain stopping power is utilizing the Bethe formula. The latest version of this method, including correction factors, is given as [5]

$$S = \frac{4\pi Z_1^2 Z_2 e^4}{mv^2} n \left\{ B_0 - \frac{C(\beta)}{Z_2} + Z_1 L_1(\beta) + Z_1^2 L_2(\beta) + \frac{1}{2}(G(M_1,\beta) - \delta(\beta)) \right\}$$

Where $B_0 = \ln\left(\frac{2mc^2\beta^2\gamma^2}{I}\right) - \beta^2$ , $e$ is the electronic charge, $n$ is the density of the atoms in the target, $m$ is mass of electron, $c$ is the speed of light, $I$ is mean excitation energy, $\beta = v/c$, $\gamma^2 = 1/(1-\beta^2)$, and $Z_1$ and $Z_2$ are atomic numbers of projectile and target, respectively. Starting with the second term, the above equation contains the following correction factors: Shell correction, Barkas term, Bloch term, Mott term, and Density correction. The experimental results and analytical predictions are in reasonable agreement for light ions. However, when the ion and/or target are heavy or the incident energy is low, the computed and experimental results exhibit considerable divergence.

Various semiempirical models for calculating stopping power can be found in the IAEA database online [4]. The Stopping Power and Range of Ions in Matter (SRIM) is the most used model, which calculates both of those quantities in the medium using a quantum mechanical treatment of ion-atom collision. Ziegler and Biersack [1] provided a detailed explanation of the calculation method. The accuracy of these calculations is fundamentally restricted because they consider only the influence of parameters during the fitting and optimization. Besides, the same degree of accuracy cannot be obtained for different cases [6]. As a result, having a new method that can obtain stopping power for any ions and targets combination over a wide range of energies with high enough accuracy is essential. A suitable approach to overcome these problems is using ML methods to predict unknown stopping powers using the current incomplete experimental datasets. Recently, ML methods have received great attention in various disciplines [7, 8]. ML gives a complex computer algorithm the ability to make predictions on a specific topic by using examples without programming the task. ML algorithms can be classified into supervised, unsupervised, and reinforcement learning [9]. The existence of labels in the training dataset distinguishes the two first primary classes based on the type of problems. Supervised learning uses labeled input and output data, while an unsupervised learning algorithm does not. In Reinforcement Learning, an agent learns in an interactive environment by trial and error using feedback from its actions and experiences [10].

Supervised learning is used in the majority of actual ML applications. Gaussian Process Regression (GPR) [11], Support Vector Machine (SVM) [12], Decision Trees (DTs) [12], and the Artificial Neural Network (ANN) [13] are frequently used algorithms. Yet, individual ML models have not consistently demonstrated to be superior even with larger training sets. Consequently, EML [14] has been developed to compensate for the inadequacies of individual learners.

EML is an advanced model that integrates multiple individual algorithms to enhance prediction accuracy. EML can be beneficial in numerous ways. Statistically, it can increase the size of the hypothesis space (H) and the possibility of finding the correct hypothesis even if the correct one does not exist in individual models' H space. In addition, the probability of a local minimum being problematic is less because multiple models start at different random initial parameters [15].

The most frequent and fundamental ensemble techniques are Bagging, Boosting, Voting, and Stacking. The Bagging algorithm is among the simplest and earliest methods of ELM techniques [16]. A random subset of data is used as a training dataset to train multiple similar models. The outputs of these models are given equal weight, and a majority voting is employed to determine the result. In the case of regression, the final output is generally the average of predictions. It helps to reduce the variance and enhance the robustness of the models. Random Forest (RF) [17, 18] and Bagging Regressor (BR) [19] are examples of the Bagging techniques.

Boosting is similar to Bagging, except it assigns various weights to each observation to boost a weak base learner to a strong one. Consequently, it assists in reducing variance and bias, as well as improving prediction accuracy. Some of the most noteworthy instances of this approach are Gradient Boosting (GB) [20] and Adaptive Boosting (AdB) [21].

While Bagging and Boosting use similar base models for ensemble, Voting and Stacking typically utilize uneven ones. The idea is that learning problems can be approached by using many models, each of which can learn a portion of the problem but not the entire problem space. As a result, you may create several separate learners and use them to create an intermediate prediction, one for each learned model. The fundamental difference between Voting and Stacking is how the final aggregation is accomplished. A Voting ensemble is obtained by adding the predictions of classification models together or averaging the predictions of regression models. In Stacking, a meta learner combines the intermediate predictions as input features to find the ultimate prediction. Thus, it significantly improves performance by Stacking individual models and their ensembles, achieving a model with superior prediction power [22].

In this work, an ML algorithm based on the Stacking model is developed and optimized using available, incomplete experimental datasets. The Stacking model was chosen among other available ML methods based on performance evaluation and maximum accuracy.

# 2- Materials and Methods

The training dataset consists of experimental electronic stopping power values available in IAEA [4]. This dataset includes summaries of all the experimental measurements conducted in various laboratories in the form of tables and figures since 1928 to the present. However, the presented data does not have a consistent format and cannot be used without pre-processing. The data cleaning, a very labor-intensive step for the described data, was accomplished manually and using Python scripts. Files and folders were renamed to be representative of the data they contained. Unnecessary information and unitless data points were eliminated, and files were converted to csv format. All energy and stopping power units were converted to MeV and $MeV \cdot cm^2 \cdot g^{-1}$, respectively. Table 1 describes the quantities within the 'cleaned' data. The final clean data includes 593 unique ion-target combinations in which the most measured values are those of hydrogen and helium projectiles in various media.

*Table 1:* Description of the quantities within the cleaned data

| Name | Number |
|---|---|
| Ion (Incident particle) | 44 |
| Target (Material under irradiation by the ion) | 93 |
| Sample (single stopping power for a particular energy, ion and target) | 40,044 |
| Energy range (MeV) | $10^{-5}$ - 985 |
| Stopping power range ($MeV \cdot cm^2 \cdot g^{-1}$) | 0.037 - 200,000 |

After training and evaluating several models, the performance of Stacking EML was found to be more robust and superior to the individual learners. Thus, we developed a two-level model using different examples of boosting and bagging algorithms as base and meta learners. RF is an extension of the Bagging algorithm frequently utilized in classification and regression applications. It constructs DT from several samples and uses their majority vote for classification and the average for regression [17]. The first Boosting algorithm was AdaBoost which employs sequential decision tree regressor models, with the subsequent model learning from the error of the previous model [21]. Because this technique is sensitive to outliers, particular attention to data is required. GB is another example of Boosting technique. Both AdaBoost and GB follow the same basic concept. In AdaBoost high weights are assigned to data points to identify the shortcomings of previous models, while in GB a negative gradient descent loss function is used for this purpose. As a result, the weights prevent over-fitting while simultaneously minimizing bias and remaining insensitive to missing datasets [23]. XGB is an improvised version of GB that is more generalizable, faster, quite accurate, and easily customizable [24]. The above models were imported from sklearn.ensemble library to develop our ML models [25]. The parameters of the estimator used in the Stacking EML method are optimized by cross-validated grid-search over a parameter grid [26]. Cross validation works by dividing training dataset into several random groups, keeping one as the test group, with the model retrained each time on the remainder until all the data has been used. The average of error metric of the models is utilized to create the final model. The Grid Search method simply checks all potential parameter value combinations and returns the one with the best accuracy.

Additionally, we performed feature selection on the model as a part of data cleaning process. Feature selection is a core concept in ML that has a significant impact on the performance of a model by reducing overfitting and complexity and improving accuracy. The most popular techniques of feature selection that identify irrelevant ones are filter methods, wrapper methods, and embedded methods [27].

The following features have been shown to be the most effective features on predicting electronic stopping power [28]: ion atomic mass, $A_\text{ion}$; ion atomic number, $Z_\text{ion}$; ion energy, $E_\text{ion}$ ; target (average) atomic mass, $A_\text{target}$ ; target (average) atomic number, $Z_\text{target}$; target density, $\rho_\text{target}$ ; target periodic group, $G_\text{target}$ ; target mean ionization energy, $I_\text{target}$ ; hydrogen fraction, $H$; carbon fraction, $C$. These parameters were included as input variables (features) in our study. Correlation coefficient was used to find the related features. The most useful variables should be highly correlated with the expected outcome but uncorrelated among themselves since two correlated features usually do not add any new information. In this work, filter method techniques (Pearson's correlation and Spearman's correlation) were used to investigate the linear and non-linear mutual correlations of the features with target and within themselves.

To assess the models four common error metrics for regression models were used. The definition of the evaluation metrics is provided in the following.

**Mean Absolute Error (MAE):** The MAE is calculated as the mean or average of the differences between predicted and expected values in a dataset.

$$MAE = \frac{1}{n}\sum_{i=1}^{n}\left|y_{true} - y_{predict}\right|$$

where $y_{true}$ is the expected value, $y_{predict}$ is the predicted value, and $n$ is the number of samples. MAE does not indicate the direction of the error meaning under prediction or over prediction the data [29].

**Mean Absolute Percentage Error (MAPE):** The MAPE calculates deviation of the residuals as a percentage.

$$MAPE = \frac{100}{n}\sum_{i=1}^{n}\left|\frac{y_{true} - y_{predict}}{y_{true}}\right|$$

The MAPE is the most often used measure for predicting error, it works best when the data does not include any extremes (and no zeros) [30].

**Root Mean Squared Error (RMSE):** The RMSE corresponds to the square root of the average of the squared difference between the true and predicted value.

$$RMSE = \sqrt{\frac{1}{n}\sum_{i=1}^{n}\left(y_{true} - y_{predict}\right)^2}$$

Because scale factors are effectively normalized, outliers are less likely to cause problems when calculating RMSE [31].

**R-Square ($R^2$):** $R^2$ calculates the amount of variance existing in the predicted dataset.

$$R^2 = 1 - \frac{\sum_{i=1}^{n}\left(y_{true} - y_{predict}\right)^2}{\sum_{i=1}^{n}(y_{true} - \bar{y}_{true})^2}$$

where $\bar{y}_{true}$ is the average of true value. A perfect model has an $R^2 = 1$ value, which indicates how close the data points are to the fitted line [32]. All preprocessing, data cleaning, modeling and evaluation of the models in this study were performed in Python 3.9. [33]

# 3- Results

## 3-1 Dataset analysis

ML models rely on training data to learn and generate predictions. Thus, data should be cleaned and investigated for the best outcome. The distribution of experimental data by energy, stopping power, and atomic masses $A_{ion}$ and $A_{target}$ (Figure 1) shows the most explored parameter intervals. Specifically, the majority of the data is concentrated within the 6 MeV range. Even though the maximum stopping power is about 200,000 MeV·cm$^2$·g$^{-1}$, there are few experimental measurements in this region. Similarly, light ions (up to 20 amu) and light targets appear to be favored more by the researchers. Targets with a mass ranging between 160 to 200 amu were also the subject of many studies. The lack of experimental data in particular parameter intervals will affect the model's accuracy when predicting stopping power values in those regions.

## 3-2 Features selection

Feature selection is defined as the process of identifying the features that contribute the most to the desired prediction, either automatically or manually. It is typically approached through evaluation of correlations between all identified model features. Figure 2 shows linear and non-linear correlation heatmaps that depict the relationship among several features. Each cell contains absolute values of the correlation coefficient. Features with high correlation are more dependent and provide almost the same information to the model. In addition, Figure 3 depicts the relationship between each feature and stopping power. This graph shows that $Z_{ion}$ and $A_{ion}$ have the most significant impact on stopping power values. In contrast, Group number and E have the least influence in terms of non-linear correlation and linear correlation, respectively.

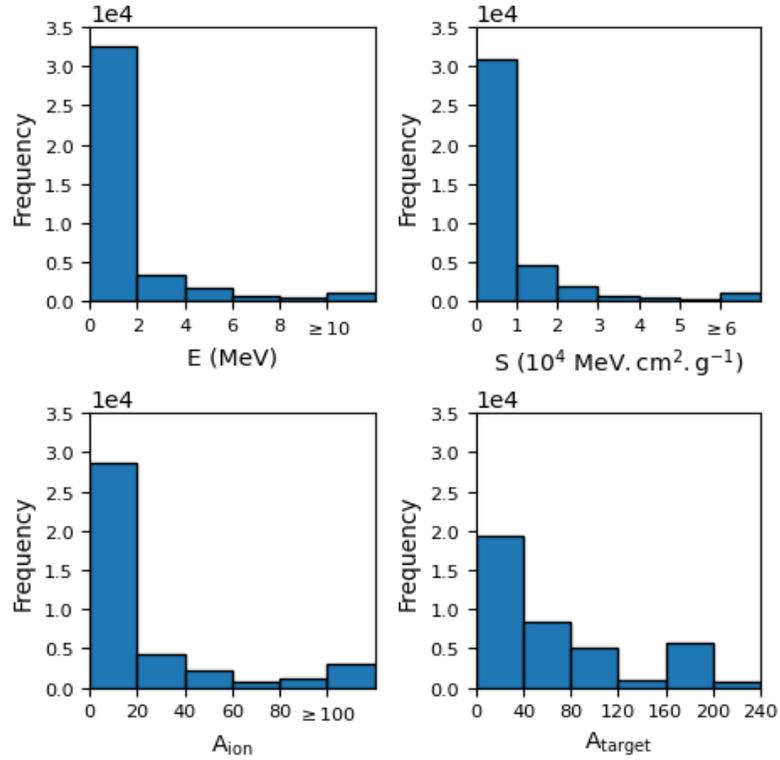

*Figure 1:* Distribution of existing experimental data by energy, stopping power, and atomic masses $A_{ion}$ and $A_{target}$.

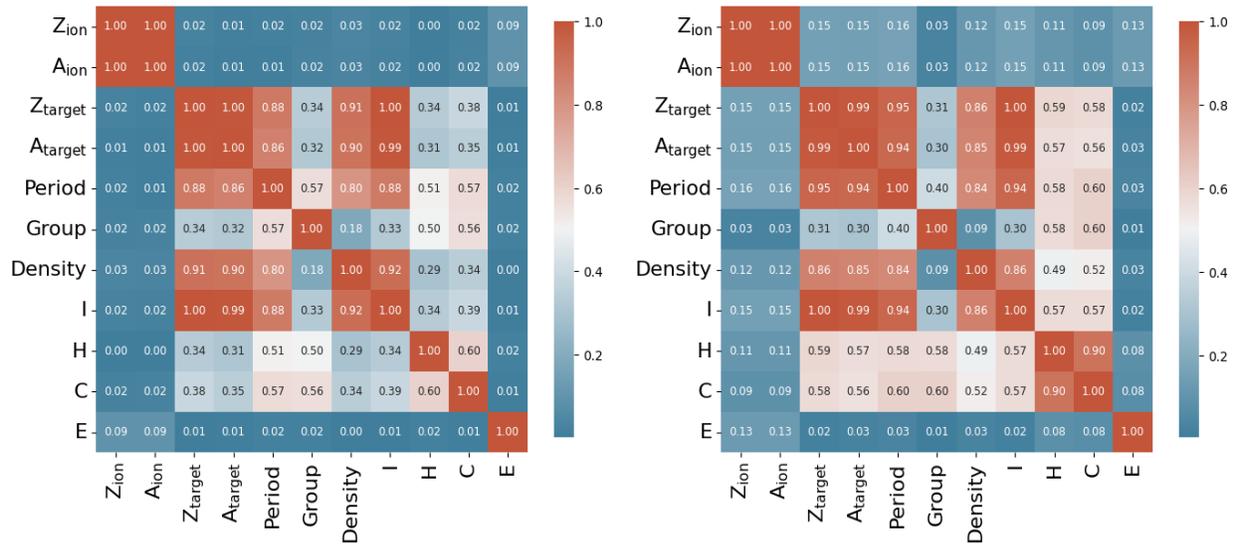

*Figure 2:* Correlation coefficients among various features, Pearson's correlation (left) and Spearman's correlation (right).

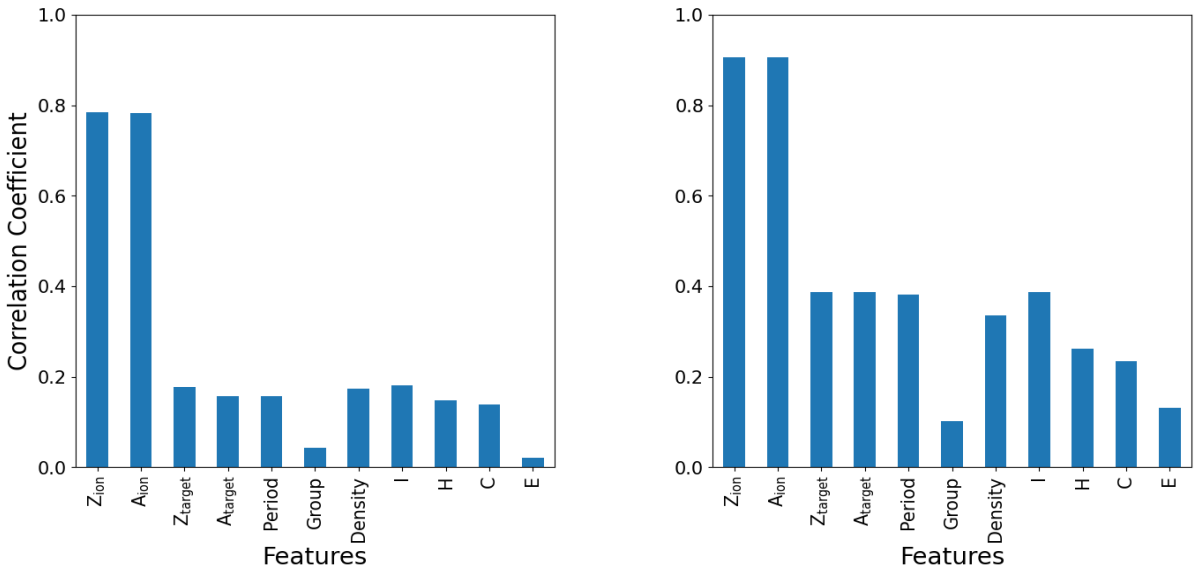

*Figure 3:* Correlation of features with stopping power, Pearson's correlation (left) and Spearman's correlation (right).

## 3-3 Models Performance

The total datasets were randomly split into a training dataset (75%) and a test dataset (25%). The test dataset was left untouched throughout the model development process to be used to evaluate the model's accuracy and reliability. The training dataset's continuous numeric variables were all rescaled to keep absolute values between 0 and 1, ensuring that numeric variables with higher absolute values were not weighted more strongly than those with lower absolute values.

Thirty different EML algorithms using various base and meta learners were employed to predict stopping power. The predictive performance of these models on training and test datasets were evaluated using different error metrics.

After ranking each models' learning accuracy in terms of MAPE on the test dataset, the top five, listed in Table 2 and Table 3, were selected. The low prediction errors across the training and testing datasets indicated that the model was neither overfitting nor underfitting the data. Combining BR, XGB, and RF as base and meta learners improved the accuracy on both training and test datasets in terms of all evaluated metrics. Low errors calculated on the training dataset, according to Table 2, imply minor bias in the model and minimal underfitting. Low testing data errors also indicated that the model was not overfitting and had low variation as shown in Table 3.

Maximum RMSE and MAE for unseen data were calculated as 1338 and 415.9 MeV·cm$^2$·g$^{-1}$, respectively which are acceptable in the wide range of stopping powers up to 200,000.00 MeV·cm$^2$·g$^{-1}$. MAPE returns error as a percentage, making the 'goodness' of the error value easier to interpret. In our study the best overall prediction error was approximately 17%.

*Table 2:* Error metrics calculated on all samples in training dataset

| Stacking Model | $R^2$ | MAE (MeV·cm$^2$·g$^{-1}$) | RMSE (MeV·cm$^2$·g$^{-1}$) | MAPE (%) |
|---|---|---|---|---|
| BR, XGB (RF) | 0.9991 | 181.0 | 511.8 | 8.750 |
| RF, XGB (BR) | 0.9985 | 220.1 | 654.8 | 10.52 |
| AdB, BR (RF) | 0.9980 | 228.6 | 769.1 | 11.58 |
| RF, GB (BR) | 0.9980 | 240.8 | 754.4 | 11.68 |
| BR, GB (RF) | 0.9977 | 247.8 | 814.1 | 12.93 |

*Table 3:* Error metrics calculated on all samples in test dataset

| Stacking Model | $R^2$ | MAE (MeV·cm$^2$·g$^{-1}$) | RMSE (MeV·cm$^2$·g$^{-1}$) | MAPE (%) |
|---|---|---|---|---|
| BR, XGB (RF) | 0.9953 | 367.1 | 1242 | 17.42 |
| RF, XGB (BR) | 0.9955 | 382.1 | 1194 | 17.21 |
| AdB, BR (RF) | 0.9948 | 402.8 | 1286 | 20.04 |
| RF, GB (BR) | 0.9944 | 415.9 | 1338 | 18.80 |
| BR, GB (RF) | 0.9946 | 409.7 | 1313 | 19.01 |

Figure 4 shows predicted values and residuals plots for all training and test datasets versus actual experimental observations. The calculated $R^2$ was used to assess each model accuracy and predictive capability. The high $R^2$ values found in this study indicated that the model was fitted to the training dataset and evaluated on the test dataset pretty accurately. As can be observed in Figure 4, the prediction model residuals are symmetrically distributed about zero and do not display any trends, pointing towards the absence of bias in the selected model. The residuals have an average value of -1.25 MeV·cm$^2$·g$^{-1}$ and 20.51 MeV·cm$^2$·g$^{-1}$ for train and test, respectively.

The model was visually assessed by investigating the error distribution. The frequency distributions of the errors calculated for each individual ion-target combinations are shown in Figure 5 along with their mean values in all datasets. Since different approaches were used to calculate mean values, these numbers should not be confused with those reported in Table 2 and Table 3. The errors calculated for individual combinations were as reasonable as those found for all samples in the datasets.

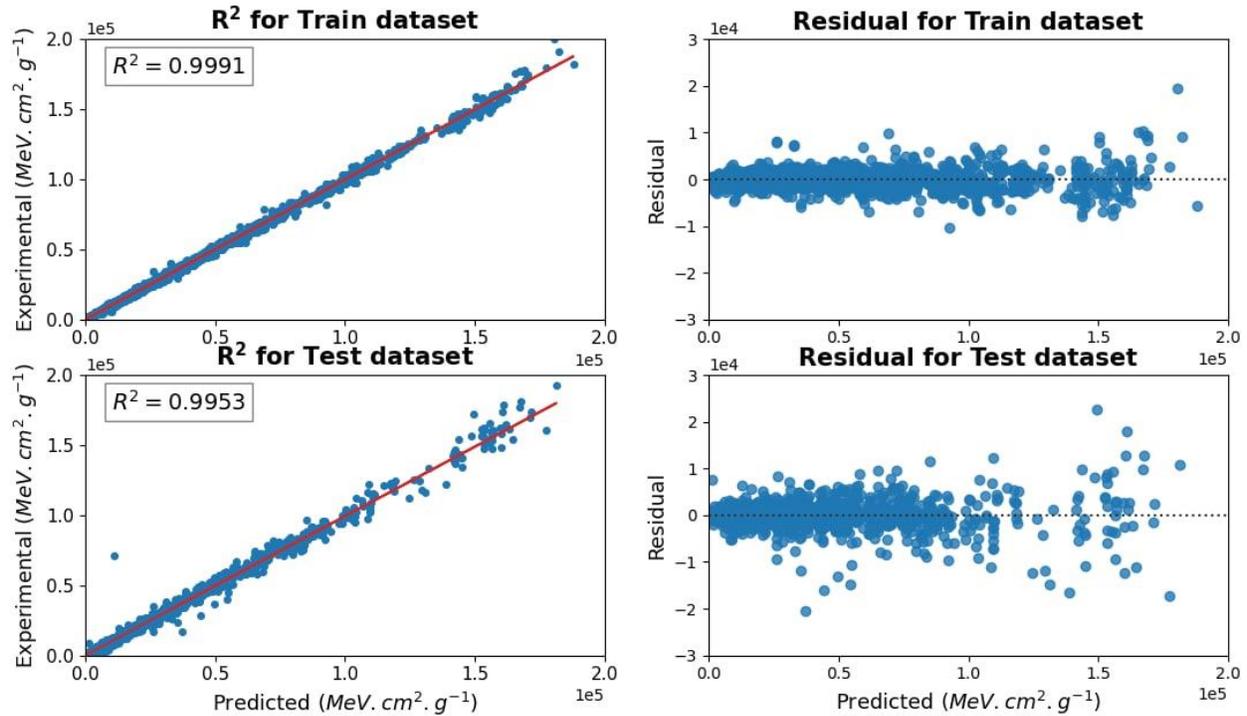

*Figure 4:* Predicted values for all samples versus experimental values with the residuals plots. Predictions are made by the model (RF, XGB (BR)) on the training data (left), and on the unseen data (right)

The benefits of removing correlated features can be in reducing overfitting, improving accuracy, and reducing the training time. As shown in Figure 2, several features were highly correlated. Exceeding a certain inter-feature correlation coefficient threshold (0.9), could mean the features provide the same information in predicting stopping power. As a result, these features were excluded from the feature set, and we retrained our models using only density, period, $Z_{ion}$, $Z_{target}$, C, and Group features which were identified as the best possible subset from the whole feature set. Comparing the obtained error metrics after removing correlated features, as shown in Table 4 and Table 5, with the values in Table 2 and Table 3 revealed a decrease only in the MAPE metric and a slight increase in the other metrics. The established underperformance of MAE and RMSE might indicate that while some of the features were correlated, they still provide information to the learner. The outperformance of MAPE can be appertained to the fact that a significant majority of the existing low stopping power data was at low energies. The bias present in the data could lead the learner to train on low stopping power regions better, leading to MAPE, which emphasizes low values, decrease.

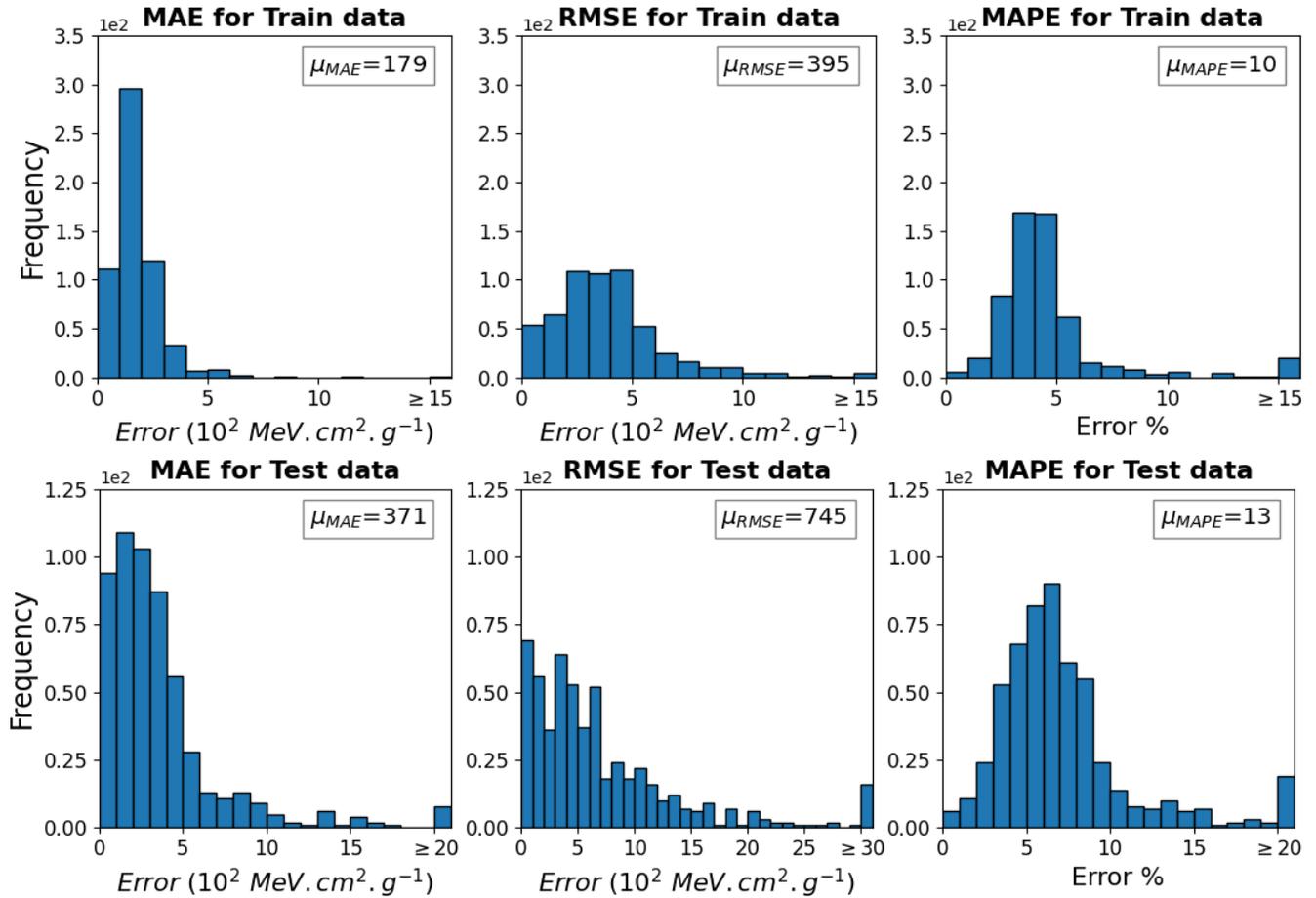

*Figure 5:* Distribution of error metrics obtained by (RF, XGB (BR)) across all ion-target combinations in the dataset. The mean error values, μ, are calculated across all combinations.

*Table 4:* Error metrics calculated on all samples in training dataset after removing highly correlated features

| Stacking Model | $R^2$ | MAE (MeV·cm$^2$·g$^{-1}$) | RMSE (MeV·cm$^2$·g$^{-1}$) | MAPE (%) |
|---|---|---|---|---|
| RF, XGB (BR) | 0.9989 | 200.8 | 556.7 | 8.179 |
| BR, XGB (RF) | 0.9991 | 182.3 | 517.9 | 8.596 |
| RF, GB (BR) | 0.9977 | 243.8 | 807.8 | 10.42 |
| BR, GB (RF) | 0.9976 | 251.2 | 831.6 | 12.22 |
| AdB, BR (RF) | 0.9978 | 228.5 | 799.8 | 9.767 |

*Table 5:* Error metrics calculated on all samples in test dataset after removing highly correlated features

| Stacking Model | $R^2$ | MAE (MeV·cm$^2$·g$^{-1}$) | RMSE (MeV·cm$^2$·g$^{-1}$) | MAPE (%) |
|---|---|---|---|---|
| RF, XGB (BR) | 0.9932 | 390.2 | 1496 | 10.34 |
| BR, XGB (RF) | 0.9933 | 381.8 | 1487 | 9.492 |
| RF, GB (BR) | 0.9915 | 429.9 | 1682 | 12.33 |
| BR, GB (RF) | 0.9913 | 448.2 | 1699 | 17.02 |
| AdB, BR (RF) | 0.9932 | 405.3 | 1496 | 11.91 |

To illustrate the best developed model performance Figure 6 shows typical outputs for some of the more important ions and targets combinations in medical physics applications, such as He_W, H_H$_2$O, H_Pb, C_Kapton, C_Air, and H_Zn (ion_target). In this figure, the original measured data, predicted values, and SRIM-2013 calculated data are plotted in red dots, blue dots, and green line, respectively, as a function of energy in the range of $10^{-3}$ to $10^2$ MeV.

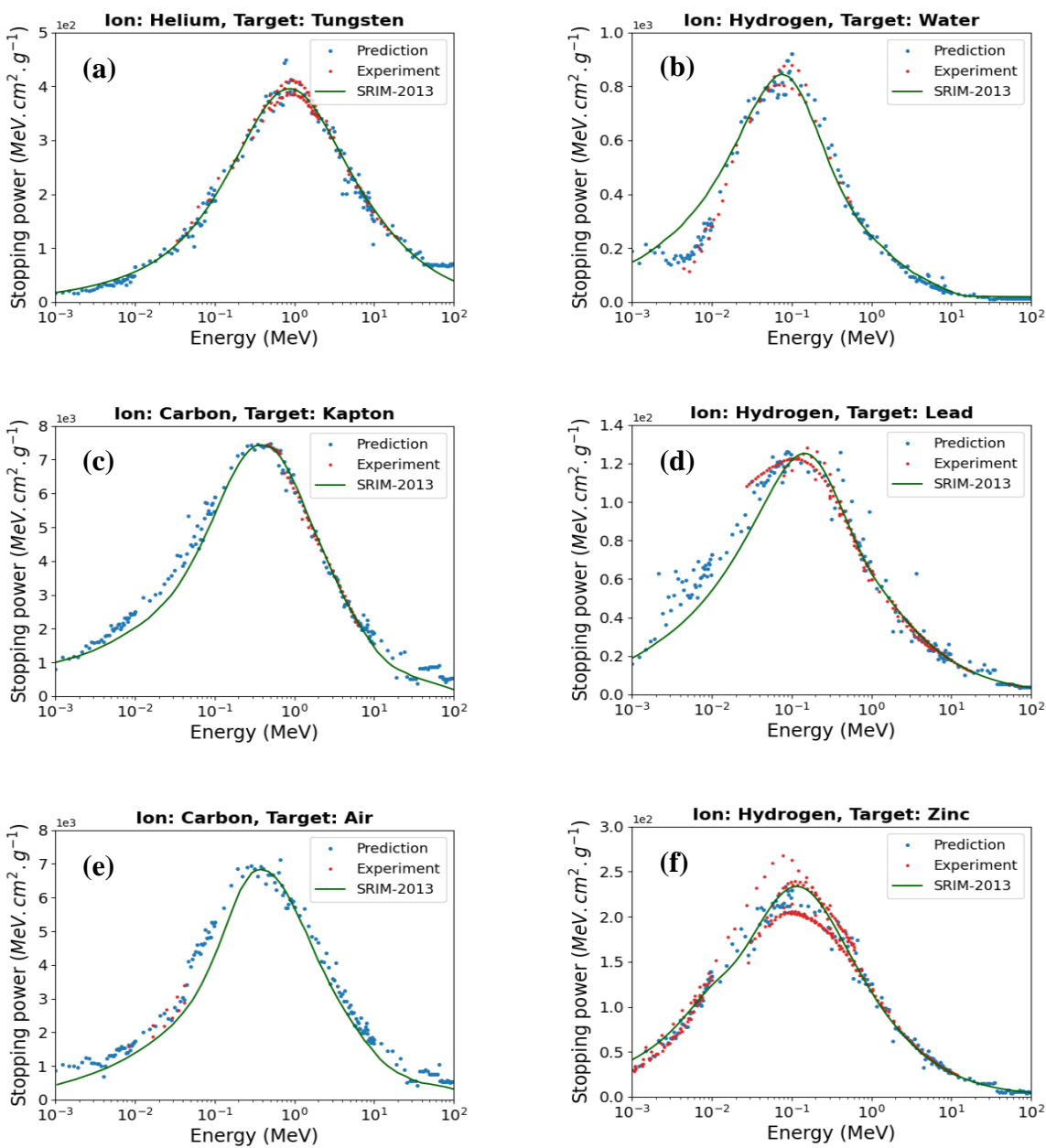

*Figure 6:* Comparison of EML predicted stopping power (blue dots) with the available measured data (red dots) and those calculated using SRIM-2013 (green line).

As shown in Figure 6(a), helium on tungsten target predicted stopping power is in good agreement with the measured data. Predicted and measured stopping powers of protons in water (Figure 6(b)) are very close, however, the SRIM model values are deviating significantly at low energies. The model generated stopping power values for polymers, heavy targets and more complex targets are also very accurate as illustrated in Figure 6 (c-e). Even in the cases where limited experimental data exist the model performs sufficiently well. Some ion-target combinations have inconsistent data points from different experiments particularly at the peak of the curve. Even in these cases the model seems to be able to make reasonable averaged predictions. An example of such a case is demonstrated in Figure 6 (e).

The model was also tested against a set of data not previously included in the training or testing dataset due to a lack of proper information on their units in the database. As can be seen in Figure 7, the model can predict the values well, indicating the correct units guessed for the data.

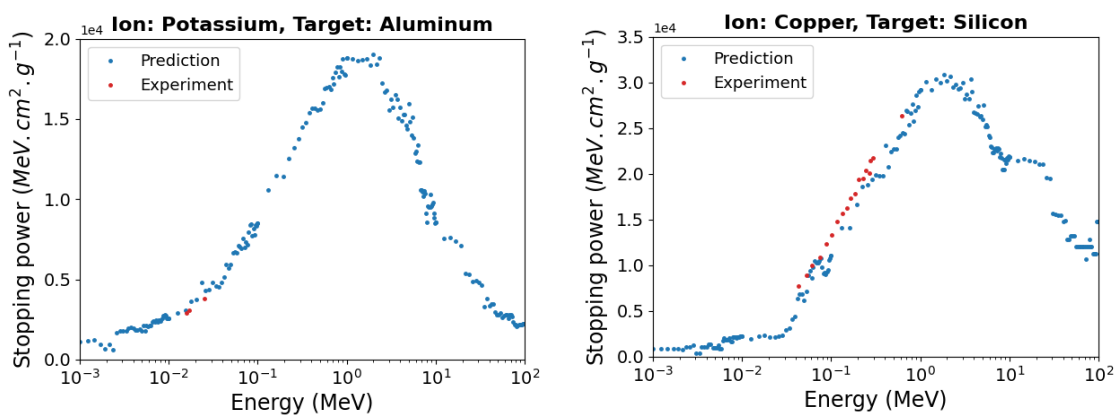

*Figure 7:* Comparison of EML predicted stopping power (blue dots) with the measured experimental data not previously included in the training or testing dataset (red dots)

## 4- Discussion

A review of stopping power publications reveals a growing interest to the subject in a variety of fields, for example, the low energy range stopping power in water and biological targets, and compound materials. Inadequate experimental data and theoretical model shortcomings motivated us to use ML as an alternative approach capable of overcoming these limitations.

Compared to theoretical databases generated by first-principles calculations or other means, training ML model with experimental data results in more accurate predictions without the need for approximations, complex functions or any other arbitrary parameters.

It is commonly understood that different model characteristics influence prediction performance. In the present study, an Stacking EML approach was utilized which, outperforms or is at least similar to a single base learner by incorporating inputs from several learning models to facilitate more accurate decisions. However, EML approach also leads to a higher computation time and complexity of the model. Two major concerns must be addressed for EML approaches. The first one relates to figuring out how to combine different base-learners. The second concern is to set appropriate parameters for each base-learners in EML. Stacking EML has the advantage of learning how to combine each of the models in an ensemble effectively. We have used five ML models, namely BR, XGB, AdB, GB, and RF, as base and meta learners to construct Stacking EML to predict electronic stopping power.

To improve the model performance, a set of features were selected from the set of all those identified in the model. Principle Component Analysis (PCA) is a common feature extraction method in data science. PCA, however, was not used in our model because the features would lose their physical meaning. We used the eleven most important features for our model based on an earlier study in which the key features have been ranked in terms of their importance.

The presence of irrelevant features in data might reduce model accuracy and cause the model to train based on irrelevant features. Our evaluations revealed the high correlation of some features. Interestingly, removing correlated features did not result in a significant model accuracy improvement, likely, because despite being correlated, they provided some valuable information to the model. Correlated features can be a problem when they do not provide any extra information and result in weaker performance specially in the case of linear models.

Each of the multiple error metric used in this study unveils unique information about the model. We have assessed the performance of our models using $R^2$, MAPE, RMSE, and MAE metrics, with the latter two being more useful for comparing the performance of different regression models. Because MAE does not identify the relative size of the error, it is not easy to distinguish between small and large errors. To deal with this problem, MAE can be used in combination with other metrics. MAPE suffers from the same limitation as MAE in that it underestimates the impact of large errors. Smaller actual values cause the relative error to worsen for the same prediction error. Because MAPE can be stated as a percentage, it offers the advantage of scale independence and interpretability, making it understandable to a broad audience. Other metrics that are not stated in relative terms or as percentages, on the other hand, frequently need subject expertise and context to appreciate the meaning of their numerical values. Larger values have more contributions to RMSE according to its definition. This enables us to use RMSE and MAE comparison as a tool to identify large, infrequent errors. The higher the difference, the greater the probability of having inconsistent predictions. RMSE has the same units as the predicted values which makes quantification of the errors possible. RMSE can be used if an application necessitates avoiding infrequent large errors in predictions. Alternatively, MAE is more informative if one wants an error metric that evaluates all errors equally and returns a more interpretable error value. MAE does not put as much weight as RMSE on the larger values and therefore it does not capture them as well as RMSE. Calculated $R^2$ on the other hand explains how well the model predicts the variance. When $R^2$ is closer to one, the predicted data matches real data more closely.

Parfitt et al. [28] have utilized an RF model to predict stopping powers. While the majority of the dataset comes from the same source, our study used approximately 17% more experimental data for model training. By comparing our results with this study, we have found our method to achieve a statistically significant improvement in all the error metric values. The Stacking EML methods used in our study generally outperform individual models; additionally, employing 5-fold grid-search (GridSearchCV) to find optimal hyperparameters also improves model accuracy.

The general feasibility of ML approaches was illustrated by comparing the predicted and measured data. Figure 6 shows that the model prediction matches the experimental measurements with high accuracy and minimal noise and discontinuities. A visualization of the data collected from simulations and those produced using the prediction model reveals a good match. SRIM, which is based on curve fitting, may be superior in areas where data is available, but is inadequate for the region where data is missing. Ideally, the data sets used for training and testing should be described in detail and made available to the public; therefore, we made all the cleaned data available on Supplementary online material.

# 5- Conclusions

The limited availability of the measured and accurately theoretically computed electronic stopping power data hinders the reliability of the related calculations in a large number of applications. We demonstrated that such a data gap can be effectively filled with machine learning approaches.

A set of 40,044 experimental datapoints, currently available from the IAEA website was used to train and test multiple ML models, both individually and in combination as ensemble models. Eleven model features were selected, based on the ranking in terms of their importance, and mutual correlations. Thirty different Stacking EML algorithms with various base and meta learners were employed to predict stopping power. Five ML models, namely BR, XGB, AdB, GB, and RF, were selected as base and meta learners to construct the final Stacking EML.

The predictive performance of these models on training and test datasets were evaluated using several error metrics, such as $R^2$, MAPE, RMSE, and MAE. Each of the error metrics unveiled unique information about the model. For ion-target combinations and energy ranges in which the experimental data exists, the predictions of our trained model outperformed any existing theoretical models, fitting the data with the average $R^2$=0.9985. The trained model also demonstrated outstanding performance ($R^2$=0.9955) with respect to the test data reserved for the model evaluation. The developed Stacking EML model can predict the stopping power for any arbitrary ion-target combination in any desired energy range. The model can serve as a universal tool for generation of the stopping power data, generally, regardless of the availability of experimental data or reliable theoretical equations in any parameter range of interest.

**Conflict of Interest:** The authors have no relevant conflicts of interest to disclose

# References

1.  Ziegler, J.F., M.D. Ziegler, and J.P. Biersack, *SRIM–The stopping and range of ions in matter (2010).* Nuclear Instruments and Methods in Physics Research Section B: Beam Interactions with Materials and Atoms, 2010. **268**(11-12): p. 1818-1823.

2.  Council, N.R., *Health effects of exposure to low levels of ionizing radiation: BEIR V.* 1990.

3.  Thomas, D.J., *ICRU report 85: fundamental quantities and units for ionizing radiation.* 2012, Oxford University Press.

4.  Paul, H., *Stopping power of matter for ions graphs, data, comments and programs (2015).* URL https://nds. iaea. org/stopping.

5.  Tai, H., et al., *Comparison of stopping power and range databases for radiation transport study.* 1997.

6.  Paul, H. and D. Sánchez-Parcerisa, *A critical overview of recent stopping power programs for positive ions in solid elements.* Nuclear Instruments and Methods in Physics Research Section B: Beam Interactions with Materials and Atoms, 2013. **312**: p. 110-117.

7.  Luo, Y., S. Chen, and G. Valdes, *Machine learning for radiation outcome modeling and prediction.* Medical Physics, 2020. **47**(5): p. e178-e184.

8.  Lam, D., et al., *Predicting gamma passing rates for portal dosimetry-based IMRT QA using machine learning.* Medical physics, 2019. **46**(10): p. 4666-4675.

9.  Burkov, A., *The hundred-page machine learning book.* Vol. 1. 2019: Andriy Burkov Quebec City, QC, Canada.

10. Sutton, R.S. and A.G. Barto, *Reinforcement learning: An introduction.* 2018: MIT press.

11. Shi, J.Q. and T. Choi, *Gaussian process regression analysis for functional data.* 2011: CRC Press.

12. Somvanshi, M., et al. *A review of machine learning techniques using decision tree and support vector machine.* in *2016 international conference on computing communication control and automation (ICCUBEA).* 2016. IEEE.

13. El-Shahat, A., *Introductory chapter: artificial neural networks.* 2018: IntechOpen.

14. Zhang, C. and Y. Ma, *Ensemble machine learning: methods and applications.* 2012: Springer.

15.    Dietterich, T.G. *Ensemble methods in machine learning*. in *International workshop on multiple classifier systems*. 2000. Springer.

16.    Galar, M., et al., *A review on ensembles for the class imbalance problem: bagging-, boosting-, and hybrid-based approaches*. IEEE Transactions on Systems, Man, and Cybernetics, Part C (Applications and Reviews), 2011. **42**(4): p. 463-484.

17.    Breiman, L., *Random forests*. Machine learning, 2001. **45**(1): p. 5-32.

18.    Kusunoki, T., et al., *Evaluation of prediction and classification performances in different machine learning models for patient-specific quality assurance of head-and-neck VMAT plans*. Medical Physics, 2022. **49**(1): p. 727-741.

19.    Sharafati, A., S.B.H.S. Asadollah, and N. Al-Ansari, *Application of bagging ensemble model for predicting compressive strength of hollow concrete masonry prism*. Ain Shams Engineering Journal, 2021. **12**(4): p. 3521-3530.

20.    Bentéjac, C., A. Csörgő, and G. Martínez-Muñoz, *A comparative analysis of gradient boosting algorithms*. Artificial Intelligence Review, 2021. **54**(3): p. 1937-1967.

21.    Schapire, R.E., *The boosting approach to machine learning: An overview*. Nonlinear estimation and classification, 2003: p. 149-171.

22.    Divina, F., et al., *Stacking ensemble learning for short-term electricity consumption forecasting*. Energies, 2018. **11**(4): p. 949.

23.    Friedman, J.H., *Stochastic gradient boosting*. Computational statistics & data analysis, 2002. **38**(4): p. 367-378.

24.    Chen, T., et al., *Xgboost: extreme gradient boosting*. R package version 0.4-2, 2015. **1**(4): p. 1-4.

25.    Pedregosa, F., et al., *Scikit-learn: Machine learning in Python*. the Journal of machine Learning research, 2011. **12**: p. 2825-2830.

26.    *https://scikit-learn.org*.

27.    Bolón-Canedo, V., N. Sánchez-Maroño, and A. Alonso-Betanzos, *A review of feature selection methods on synthetic data*. Knowledge and information systems, 2013. **34**(3): p. 483-519.

28.    Parfitt, W.A. and R.B. Jackman, *Machine learning for the prediction of stopping powers*. Nuclear Instruments and Methods in Physics Research Section B: Beam Interactions with Materials and Atoms, 2020. **478**: p. 21-33.

29. Willmott, C.J. and K. Matsuura, *Advantages of the mean absolute error (MAE) over the root mean square error (RMSE) in assessing average model performance.* Climate research, 2005. **30**(1): p. 79-82.

30. Tayman, J. and D.A. Swanson, *On the validity of MAPE as a measure of population forecast accuracy.* Population Research and Policy Review, 1999. **18**(4): p. 299-322.

31. Chai, T. and R.R. Draxler, *Root mean square error (RMSE) or mean absolute error (MAE)?–Arguments against avoiding RMSE in the literature.* Geoscientific model development, 2014. **7**(3): p. 1247-1250.

32. Alexander, D.L., A. Tropsha, and D.A. Winkler, *Beware of R 2: simple, unambiguous assessment of the prediction accuracy of QSAR and QSPR models.* Journal of chemical information and modeling, 2015. **55**(7): p. 1316-1322.

33. *https://www.python.org/.*